\documentclass[manuscript]{acmart}
\pdfoutput=1

\usepackage[english]{babel}
\usepackage[utf8x]{inputenc}
\usepackage[T1]{fontenc}
\usepackage{amsmath}
\usepackage{graphicx}
\usepackage{booktabs}
\usepackage{csquotes}
\usepackage{enumitem}
\usepackage{hyperref}
\usepackage{xcolor}
\usepackage{rotating}
\usepackage{microtype}        
\usepackage{ccicons}          

\definecolor{orange}{HTML}{1b9e77}
\definecolor{green}{HTML}{d95f02}
\definecolor{gold}{HTML}{e8bc59}
\definecolor{blue}{HTML}{314283}
\definecolor{yellow}{HTML}{bc9437}

\PrerenderUnicode{ü}
\newcommand{\para}[1]{\vspace{3pt}\textit{#1~}}

\copyrightyear{2022} 
\acmYear{2022} 
\setcopyright{acmlicensed}\acmConference[FAccT '22]{2022 ACM Conference on Fairness, Accountability, and Transparency}{June 21--24, 2022}{Seoul, Republic of Korea}
\acmBooktitle{2022 ACM Conference on Fairness, Accountability, and Transparency (FAccT '22), June 21--24, 2022, Seoul, Republic of Korea}
\acmPrice{15.00}
\acmDOI{10.1145/3531146.3533097}
\acmISBN{978-1-4503-9352-2/22/06}

\begin{document}

\author{Maurice Jakesch}
\affiliation{%
  \institution{Cornell University}
  \city{New York}
  \state{NY}
  \country{USA}
}
\email{mpj32@cornell.edu}
\author{Zana Buçinca}
\affiliation{%
  \institution{Harvard University}
  \city{Boston}
  \state{MA}
  \country{USA}
}
\author{Saleema Amershi}
\affiliation{%
  \institution{Microsoft Research}
  \city{Redmond}
  \state{WA}
  \country{USA}
}
\author{Alexandra Olteanu}
\affiliation{%
  \institution{Microsoft Research}
  \city{Montreal}
  \state{QC}
  \country{Canada}
}


\title{How Different Groups Prioritize Ethical Values for Responsible AI}


\begin{abstract}
Private companies, public sector organizations, and academic groups have outlined ethical values they consider important for responsible artificial intelligence technologies. While their recommendations converge on a set of central values, little is known about the values a more representative public would find important for the AI technologies they interact with and might be affected by. We conducted a survey examining how individuals perceive and prioritize responsible AI values across three groups: a representative sample of the US population (N=743), a sample of crowdworkers (N=755), and a sample of AI practitioners (N=175). Our results empirically confirm a common concern: AI practitioners' value priorities differ from those of the general public. Compared to the US-representative sample, AI practitioners appear to consider responsible AI values as less important and emphasize a different set of values. In contrast, self-identified women and black respondents found responsible AI values more important than other groups. Surprisingly, more liberal-leaning participants, rather than participants reporting experiences with discrimination, were more likely to prioritize fairness than other groups.
Our findings highlight the importance of paying attention to who gets to define responsible AI. 

\end{abstract}

\maketitle

\section{Introduction}

Advances in artificial intelligence (AI) have the potential to benefit people and society, but they also raise ethical challenges and concerns about possible adverse impacts~\cite{montreal}. Being prone to errors and biases, AI systems may harm people~\cite{awad2018moral} for instance by reinforcing stereotypes~\cite{Blodgett_power_2020} or by increasing social inequality~\cite{eubanks2018automating}. While the larger consequences of AI can be difficult to anticipate~\cite{Boyarskaya_Olteanu_Crawford_2020}, systems developed with broader human and societal values in mind stand a better chance of preserving these values~\cite{raji2020closing, friedman1996value, agre1997computation}. To support the development of socially beneficial AI technologies, several private companies, public sector organizations, and academic groups have published ethics guidelines with values they consider important for responsible AI~\cite{Jobin_Ienca_Vayena_2019}. 

These AI ethics guidelines have been found to largely converge on five central values~\cite{Jobin_Ienca_Vayena_2019}: transparency, fairness, safety, accountability, and privacy. 
But these values may differ from what a broader and more representative population would consider important for the AI technologies they interact with. 
While prior work has shown that value preferences depend on peoples' backgrounds and personal experiences~\cite{sep-scheler, intemann201025}, AI technologies are often developed by relatively homogeneous and demographically skewed subsets of the population~\cite{landivar2013disparities, crawford2016artificial, house2016preparing}. 
Given the lack of reliable data on other groups' priorities for responsible AI, practitioners may unknowingly encode their own biases and assumptions into their concept and operationalization of responsible AI~\cite{martin2019ethical, raji2020closing}. 

In this work, we present the results of a survey we developed, validated, and fielded to elicit peoples' value priorities for responsible AI. Drawing on the traditions of empirical ethics~\cite{musschenga2005empirical, Dunn_Sheehan_Hope_Parker_2012} and value elicitation research~\cite{fischhoff1991value, schwartz2007basic}, our survey asks participants about the perceived importance of a set of 12 responsible AI values both in general and in specific deployment scenarios. To increase robustness, respondents assessed values from three perspectives: value selection, contextual assessment of values, and comparative prioritization of values (detailed in \S\ref{subsec:design}).  

We administered this survey to three different populations. 
We analyzed how value priorities of a US census-representative sample (N=743), a crowdworker sample (N=755), and an AI practitioner sample (N=175) vary by deployment scenario and individuals' backgrounds and experiences. We surveyed the value priorities of AI practitioners as they are often the ones making decisions about the AI technologies that are being developed, and compared their preferences to those of a more representative sample. We also consulted crowdworkers as they are already involved in producing data that AI systems are evaluated on to explore the feasibility of involving them in the ethical assessment of AI systems as well.

Our results provide evidence that responsible AI values are perceived and prioritized differently by different groups. AI practitioners, on average, rated responsible AI values less important than other groups. At the same time, AI practitioners prioritized fairness more often than participants from the US-census representative sample who emphasized safety, privacy, and performance. 
We also find differences in value priorities along demographic lines. For example, women and black respondents evaluated responsible AI values as more important than other groups.
We observed the most disagreement in how people traded-off fairness with performance.
Surprisingly, participants reporting past experiences of discrimination did not prioritize fairness more than others, but liberal-leaning participants prioritized fairness more than conservative-leaning participants.

Our results highlight the need for AI practitioners to contextualize and probe their ethical intuitions and assumptions.  
The empirical approach to AI ethics explored in this study can help to increase the context sensitivity of the responsible AI development process. 
However, as we elaborate in the discussion, opinion research can inform ethical decision-making, but cannot replace sound ethical reasoning. 

\section{Background}
Our study draws on prior work on responsible AI, value sensitive design~\cite{friedman1996value}, empirical ethics~\cite{musschenga2005empirical}, value elicitation~\cite{fischhoff1991value, schwartz2007basic}, and standpoint theory~\cite{intemann201025}.

\subsection{AI ethics guidelines and value-sensitive design}
Science and technology studies theorize that computing technologies incorporate a tacit understanding of human nature~\cite{winograd1986understanding}. Algorithms are described as value-laden artifacts~\cite{martin2019ethical} that encode developer assumptions, including ethical and political values~\cite{raji2020closing}. From this perspective, a product team that decides to maximize the chance that a disease detection system will recognize a disease at the cost of increasing false alarms prioritizes certain values over others. Past work has shown that machine learning development and research often narrowly focus on technical values such as accuracy, efficiency, and generalization~\cite{Birhane_Kalluri_Card_Agnew_Dotan_Bao_2021, nanayakkara2021unpacking}. In contrast, proponents of value-sensitive design~\cite{friedman1996value, friedman1996bias}, reflective design~\cite{sengers2005reflective}, and critical technical practice~\cite{agre1997computation} advocate that AI systems should be designed with broader human and societal values in mind.

What values developers of responsible AI systems should emphasize remains a key question. Some argue these values should be naturally embedded in an organization's culture~\cite{raji2020closing}. Several organizations have also published guidelines describing what values they believe AI systems should embody. \citet{Jobin_Ienca_Vayena_2019} found these guidelines to converge around central values, but differ in how they construe these values and concepts. 
Critics note that reliable methods to translate values into practice are often missing~\cite{raji2020closing, mittelstadt2019ai}. Some also argue that statements of high-level values and principles are too ambiguous and may gain consensus simply by masking the complexity and contending interpretations of ethical concepts~\cite{whittlestone2019role}. For example, people may agree on the importance of fairness, but ``fairness'' in and by itself has little to say about what is fair and why~\cite{binns2018fairness}. 

Our study validates and contextualizes value priorities outlined in AI ethics guidelines. To date, there is little empirical data on values a broader and more representative public finds important for the AI technologies they interact with.
Our empirical approach to AI ethics probes for possible blind spots in AI practitioners' and researchers' assumptions. 

\subsection{Empirical studies of human values and AI ethics}
Eliciting people's values is a central pursuit in the social sciences~\cite{fischhoff1991value}. Economists explain choices in the marketplace based on value theory, sociologists seek to understand which values are held by a community and how they change. Psychologists use value elicitation for therapy and counsel, and empirical ethicists enhance the context-sensitivity of their arguments by combining social scientific methods with ethical reasoning~\cite{musschenga2005empirical}. While drawing normative conclusions from empirical results is difficult, empirical data on ethical preferences can inform decision making~\cite{musschenga2005empirical}. 

Several studies have examined people's ethical intuitions concerning AI technologies. In the ``moral machine'' experiment, \citet{awad2018moral} generated a variety of moral dilemmas a self-driving car might find itself in and ask participants which course of action they recommend. They report significant cross-cultural differences in ethical preferences correlated with modern institutions and cultural traits. 
\citet{hidalgo2021humans} explored how people judge humans and machines differently when they make mistakes. They found that people tend to forgive machines more in scenarios with high intentionality. Similarly, \citet{malle2015sacrifice} compared how people apply moral norms to humans versus robots. Most related to the empirical study of responsible AI values,  \citet{Saxena_Huang_DeFilippis_Radanovic_Parkes_Liu_2019} have compared public perceptions of different fairness paradigms. Similarly, \citet {grgic2018human} and \citet{pierson2017demographics} have studied which features people find fair to include in a prediction algorithm. 
They found substantial disagreement among participants~\cite{grgic2018human}, with e.g., women being less likely to include gender as a feature in a course recommendation algorithm if this might result in female students seeing fewer recommendations for science courses~\cite{pierson2017demographics}.

Going beyond previous work, we develop a responsible AI value survey to explore what values people find most important for responsible AI. Where previous studies have elicited preferences concerning specific technical implementations with convenience samples, we provide a first high-level perspective on a representative public's priorities for the AI system they interact with and might be affected by.

\subsection{The impact of individual background \& of context on how values are prioritized}
Feminist empiricists and standpoint theorists argue that knowledge is achieved from a particular standpoint~\cite{wylie2003standpoint} and that social location systematically influences our experiences and decisions~\cite{intemann201025}. They hold that homogeneous communities are prone to false consensus effects~\cite{ross1977false} where individuals believe that the collective opinion of their own group matches that of the larger population. In homogeneous communities, inaccurate assumptions or biases can be hard to recognize and correct~\cite{intemann201025, Boyarskaya_Olteanu_Crawford_2020}.
In communities comprised of individuals with diverse values and experiences, however, how assumptions influence reasoning becomes more visible~\cite{intemann201025, longino2020science, rolin2006bias}. Including historically underrepresented groups, in particular, may lead to rigorous critical reflection as their experiences may facilitate the identification of problematic background assumptions~\cite{intemann201025}. 

Demographics and experiences not only affect background assumptions~\cite{dobbe2018broader}, but also shape people's values and ethical preferences~\cite{fumagalli2010gender, graham2016cultural}. 
Rather than stemming from overarching belief systems, values often arise through particular social practices in a specific context~\cite{macintyre1981nature}.
As such, ethical intuition is contextual and socially situated~\cite{sep-scheler}. 
For instance, what's fair to some people may seem unfair to others~\cite{lee2017algorithmic}, and some people value privacy and autonomy more than others~\cite{whittlestone2019role}. 
The population of AI practitioners is demographically skewed~\cite{landivar2013disparities, crawford2016artificial, house2016preparing} with e.g., women and black people being underrepresented~\cite{dillon2015state}.
With their specific demographics and experiences, AI practitioners may bring their own preferences to what it means for AI to be ``responsible'' or ``ethical'', such as a bias towards deployment~\cite{kaur2020interpreting}. Responsible AI technologies developed within homogeneous communities may fail to account for the experiences and needs of various groups, so it remains crucial to scrutinize who gets to define AI ethics~\cite{jobin2021ai}.

By surveying representative population samples about their priorities for responsible AI, we seek to validate the value prioritization in AI ethics frameworks. We explore the social relativity of responsible AI values to provide grounds for more critical reflection about possibly inaccurate assumptions and false consensus effects. 
\section{Methods}

To study how people perceive and prioritize responsible AI values, we combine instruments from value elicitation research~\cite{fischhoff1991value} with the concepts and principles found in AI ethics guidelines~\cite{Jobin_Ienca_Vayena_2019}. We fielded an iteratively developed online survey with 743 census-representative participants, 755 crowd workers, and 175 AI practitioners. 

\subsection{Survey development}

\begin{figure*}
  \begin{center}
    \includegraphics[width=0.99\textwidth]{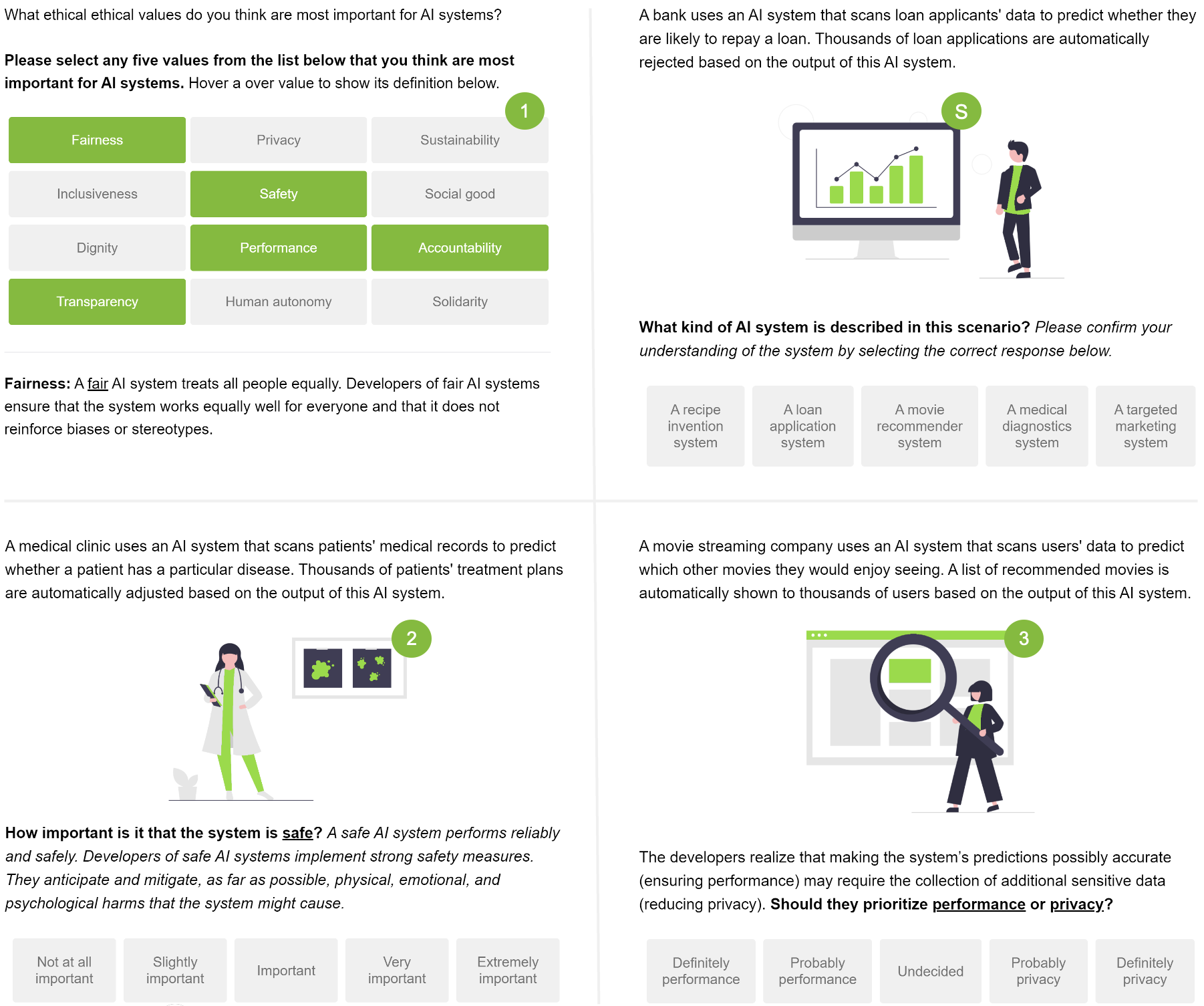}
  \end{center}
\caption{Overview of the main survey components. Participants first completed a value selection task {\bf (1)}. After confirming the understanding of the respective deployment scenario {\bf (S)}, they evaluated how the importance of values in context {\bf (2)}. Finally, participants indicated how they would prioritize values when they are in conflict  {\bf (3)}. 
} 
\label{fig:procedure}
\end{figure*}

We adapted the Schwartz Value Survey~\cite{schwartz1992universals, schwartz1994there} to apply it to responsible AI values. 
The Schwartz Value Survey has been used to study individual and intercultural differences in general human values in over 60 countries~\cite{schwartz2007basic}. 
Based on an inventory of human values, the Schwartz Value Survey asks respondents to self-report which values are most important to them. Respondents rate the importance of each value on a Likert scale while explanations for each value are shown. 

\para{Selecting and explaining responsible AI values. }
To adapt the Schwartz Value Survey to the study of AI ethics, we constructed an inventory of responsible AI values.
The responsible AI values we chose for our survey are based on a review of published AI ethics guidelines. We drew on work by \citet{Jobin_Ienca_Vayena_2019} finding that AI ethics guidelines commonly refer to transparency, justice \& fairness, non-maleficence, accountability, privacy, beneficence, freedom \& autonomy, trust, and dignity. To this list, we added system performance, as it is a central value in AI research and development~\cite{Birhane_Kalluri_Card_Agnew_Dotan_Bao_2021} that is often used to compare AI models and to make deployment decisions. 

As responsible AI values are abstract and participants may not easily understand how they apply in the context of AI technologies~\cite{cave2018portrayals}, we provided additional explanations.
To formulate  explanations for each value, we drew again on existing AI ethics guidelines, including Microsoft's responsible AI principles~\cite{microsoft}, Google's AI Principles~\cite{google}, the Montreal Declaration for the Responsible Development of Artificial Intelligence~\cite{montreal}, the Deloitte AI ethics guide~\cite{deloitte}, IBM's Principles for Trust and Transparency~\cite{ibm}, and the EU's Ethics guidelines for trustworthy AI~\cite{eu}.

We tested and iterated on different explanations of responsible AI values in four crowdsourcing pilot studies (N\textsubscript{1}=40, N\textsubscript{2}=80, N\textsubscript{3}=40, N\textsubscript{4}=160). Each pilot asked participants whether they understood an explanation through both Likert scales and open-ended responses. Based on the pilot results, we substituted ``non-maleficence'' with ``safety'' and ``beneficence'' with ``social good,'' as the former were not well-understood by participants. We also explicitly referred to ``\emph{human} autonomy'' to avoid confusion with autonomous cars and robots. Finally, we did not include ``trust'' as it appeared overly general and overlapped with other values such as transparency and accountability. 

We phrased the explanations in simple, non-technical language, all following the same structure. Each explanation starts with a sentence describing what a system embodying the value would do, followed by an example of steps developers might take to realize a value, e.g.: ``{\em An AI system that respects people's autonomy avoids reducing their agency. Developers of autonomy-preserving AI systems ensure, as far as possible, that the system provides choices to people and preserves or increases their control over their lives.}'' By complementing a general definition with specific operationalizations of a value, the framing provides a tangible understanding of the value while maintaining a degree of generality. See Appendix~\ref{appendix:RAI-values} for a complete list of the explanations we used in our survey. 

\para{Identifying pairs of possibly conflicting responsible AI values.}
In addition to assessments of values themselves, we asked participants about their preferences in cases of conflicting values~\cite{barocas2017engaging}. For example, ensuring fairness might require collecting additional sensitive data, potentially diminishing privacy. 
To identify value conflicts, we searched for mentions of conflicts in the literature for each pair of values in the responsible AI value inventory. We found prior discussions of trade-offs between privacy \& performance~\cite{bagdasaryan2019differential, shokri2015privacy}, 
fairness \& privacy~\cite{bagdasaryan2019differential, ekstrand2018privacy}, 
fairness \& performance~\cite{corbett2017algorithmic, pleiss2017fairness}, 
safety \& transparency~\cite{hua2021increasing, meijer2014reconciling, cappelli2010transparency}, and autonomy \& safety~\cite{livingstone2011risks}. We combined the value explanations developed above to introduce the conflicts to participants, e.g. ``{\em The developers realize that minimizing the collection of sensitive data (ensuring privacy) may make the system's predictions less accurate (reducing performance). Should they prioritize privacy or performance?}''

\para{Constructing hypothetical AI deployment scenarios.}
We used hypothetical scenarios to make value assessments more tangible and to elicit judgments in specific contexts. 
We produced four hypothetical deployment settings validated through two pilot studies (N\textsubscript{1}=180, N\textsubscript{2}=160). 
To design these scenarios, we selected 25 AI systems people may have encountered in everyday settings starting with a list of general AI use cases~\cite{aimultiple}. We developed short explanations of these use cases and asked pilot participants whether they found them understandable and relatable. Based on the pilot results, we further refined the scenarios and kept only the 10 scenarios that were most easily understood by pilot participants. The second pilot then asked participants which scenarios they understood best and whether the AI system's decisions were highly consequential. Based on the responses, we selected two well-understood high-stake and low-stake scenarios for the study:

\begin{enumerate}[label=(\alph*)]
\item Medical: An AI system used by a medical clinic to predict whether a patient has a disease (high-stake)
  \item Banking: An AI system used by a bank to predict whether an applicant will repay a loan (high-stake)
  \item Marketing: An AI system used by a marketing company to match ads to viewers (low-stake)
  \item Streaming: An AI system used by a streaming company to recommend movies to users (low-stake)
\end{enumerate}
Each scenario states the entity controlling the AI system and the type of data the system is using. It then elaborates what predictions are being made and what actions are being taken based on the prediction, e.g.: ``A medical clinic uses an AI system that scans patients' medical records to predict whether a patient has a particular disease. Thousands of patients' treatment plans are automatically adjusted based on the output of this AI system.'' The full list of scenarios is included in  Appendix~\ref{appendix:scenarios}.

\subsection{Survey procedure} 
\label{subsec:design}

After providing informed consent, participants received a high-level introduction both covering the general goals of AI and noting the complex decision-making involved in the AI system development beyond technical challenges (see Appendix~\ref{appendix:intro}). 
Figure \ref{fig:procedure} illustrates the subsequent survey steps which combined three value elicitation tasks: 
(1) {\em value selection}---select five responsible AI values (out of the 12) that are deemed most important in general, 
(2) {\em contextual assessment}---evaluate the perceived importance of seven central responsible AI values (transparency, fairness, safety, accountability, privacy, autonomy, and performance) in a specific deployment setting, and 
(3) {\em comparative assessment}---recommend what product teams should do when values are in conflict. 

Participants selected the five most important values for AI systems in general, with explanations displayed when a value was hovered over. They then read the first scenario and confirmed their understanding of the deployment setting. Overall, participants encountered four scenarios. In scenarios 1 and 2, participants indicated how important they thought three responsible AI values were in the given situation on a 5-point Likert scale. In scenario 3, participants evaluated one more value and then two value conflicts by indicating which value they thought should be prioritized in the given situation. Finally, they evaluated three value conflicts in the fourth and last scenario. For every rating, participants were given the option to explain their choices.

After completing the rating tasks, participants indicated their familiarity with machine learning, user research, and their personal experiences with discrimination.
We selected these experiential correlates based on the hypothesis that personal experience might inform ethical preferences~\cite{sep-scheler}. For example, user researchers may have learned to empathize with users, whereas respondents trained in ML may have better insight into the technical constraints of responsible AI.
We also asked participants to report their gender identity, age, ethnicity, political views, sector of work, and highest level of education. 
Again, these demographic correlates were selected to explore to what extent social location influences the perceived importance of responsible AI values~\cite{intemann201025}. 
For all experiential and demographic questions, participants could choose not answer. 

\subsection{Participant recruitment}
To examine how different groups assess responsible AI values, we surveyed three populations: 

\para{A US census-representative sample} (N=743\textsubscript{1}) was recruited by Qualtrics to gain insights into how the general population assesses the importance of responsible AI values. The recruitment process combined a variety of methods to minimize biases and performed stratified random sampling to match the US census along gender, age, race, region, and household income. Participant compensation was handled by Qualtrics. 

\para{A convenience US-based crowdworker sample} (N\textsubscript{2}=755) was recruited via the Clickworker crowdsourcing platform. Participants were US-based and likely previously contributed to the training of AI models by e.g., providing data labels. Each participant received USD 2.8 for a median participation time of 8 minutes. While crowdworkers are not directly involved in the AI development process, their judgments are often a key ingredient to machine learning systems. We explored whether their assessments could serve as proxies for the ethical intuition of a more representative population. 

\para{A sample of AI practitioners} (N\textsubscript{3}=175) was recruited through an open call on Twitter (N=156) and internal mailing lists (N=19) at a large tech company. Our call for participation targeted US-based participants whose work is related to AI/ML. We confirmed their background in the survey, but ultimately rely on self-reported expertise. For the internal mailing lists, we specifically targeted teams doing AI/ML related work. Participants could choose to enter a raffle to win one of five \$50 gift vouchers after study completion. AI practitioners are a relevant population that makes key decisions throughout the AI development process. We explore whether their value judgments differ from those of the more general population.

We had to work with different types of compensation due to differences in respondent type and recruitment method across samples. However, we aimed to provide roughly commensurate compensation across recruitment methods. The study was IRB approved, and we obtained informed consent from all our participants.

\subsection{Data quality control} 
To counterbalance ordering effects, the arrangement of scenarios, values, and conflict questions was randomized. In addition, the order of response options was randomly flipped per participant. For the conflict questions, we also randomized the internal order of the conflict, e.g. fairness vs. performance was inverted to performance vs. fairness. 
A pop-up window asked participants to slow down whenever they attempted to submit responses in under 3 seconds per survey page to deter spammers and inattentive participants.
The four scenario introductions throughout the survey served as attention and comprehension checks for our participants. We removed all participants that had failed more than one attention check from our analysis to increase response quality, reducing the relevant samples to N\textsubscript{1}=516, N\textsubscript{2}=607, N\textsubscript{3}=140 respectively.

\section{Results}

\subsection{What values are deemed as most important in general?}

\begin{figure*}
  \begin{center}
    \includegraphics[width=0.8\textwidth, , trim=0cm 0.4cm 0cm 0cm]{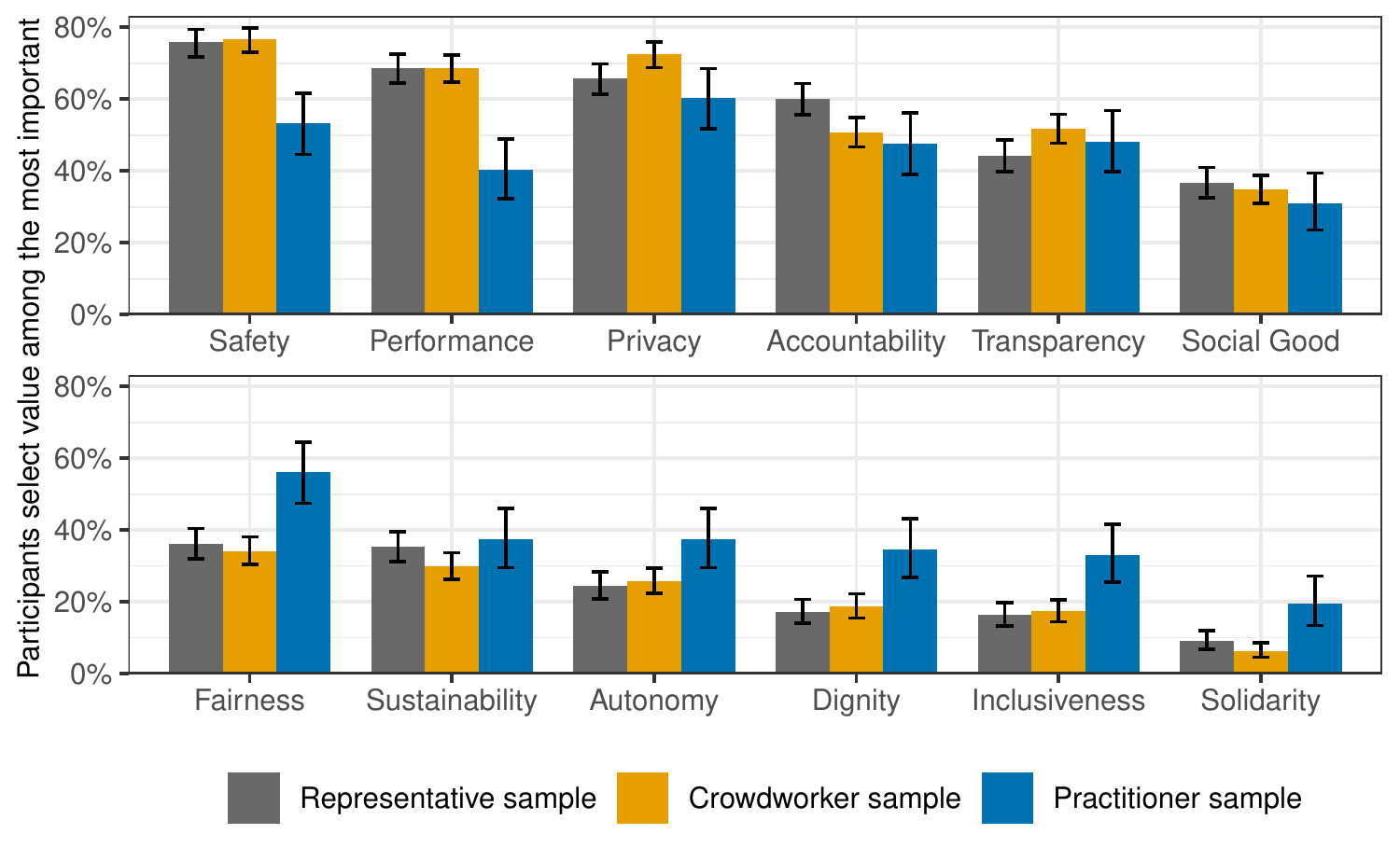}
  \end{center}
\caption{\textbf{AI practitioners' value priorities differ from those of the general public.} N\textsubscript{1}=516, N\textsubscript{\textcolor{gold}{2}}=607, N\textsubscript{\textcolor{blue}{3}}=140. 
The x-axis shows the 12 responsible AI values respondents chose from, while the y-axis indicates how often respondents selected a value among the five most important. Participants from the US-census representative sample and the crowdworker sample selected safety, performance, and privacy most often among their five most important values, while practitioners selected fairness more often.}
\label{fig:task1-sample}
\end{figure*}

In Task 1, participants selected five values they deemed most important for AI systems out of an inventory of 12 responsible AI values (Figure \ref{fig:task1-sample}). 
76\% of respondents from the US-census representative sample selected safety among the top 5 responsible AI values. Over 60\% of participants in this representative panel also selected performance, privacy, and accountability among the most important values. Respondents from the crowdworker sample selected accountability less often, but their preferences were largely consistent with those from the US-census representative sample.
AI practitioners' preferences were less focused. Compared to the US-census representative sample, practitioners selected humanist values such as fairness, inclusiveness, dignity, and solidarity more often and were less likely to select safety and performance among the most important values.

\subsection{How important are values in specific deployment scenarios?}

\begin{figure*}[t]
  \begin{center}
    \includegraphics[width=0.8\textwidth, trim=0cm 0.4cm 0cm 0cm]{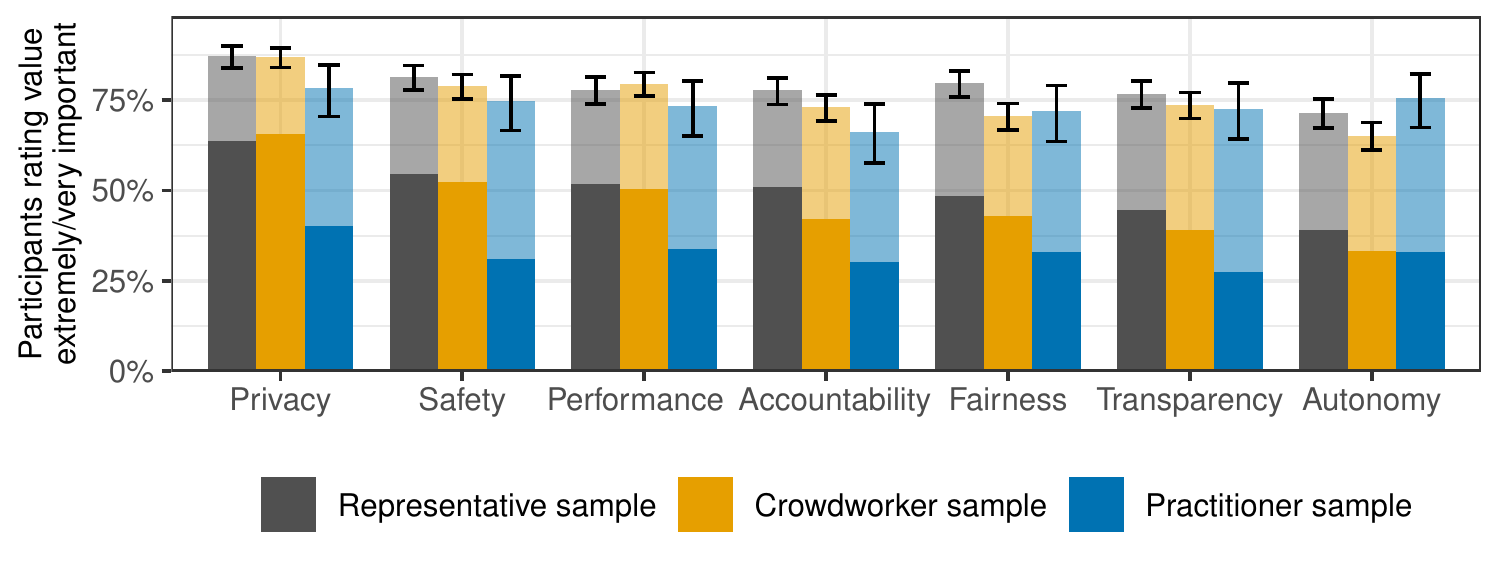}
  \end{center}
\caption{\textbf{Representative participants rated responsible AI values as more important than AI practitioners did.} N=140 to 607 ratings per bar. The x-axis shows the assessed responsible AI values and the y-axis indicates how often respondents evaluated the responsible AI value as very important (light) or extremely important (dark).}
\label{fig:task2-sample}
\end{figure*}

\begin{figure*}[t]
  \begin{center}
    \includegraphics[width=0.8\textwidth, trim=0cm 0.4cm 0cm 0cm]{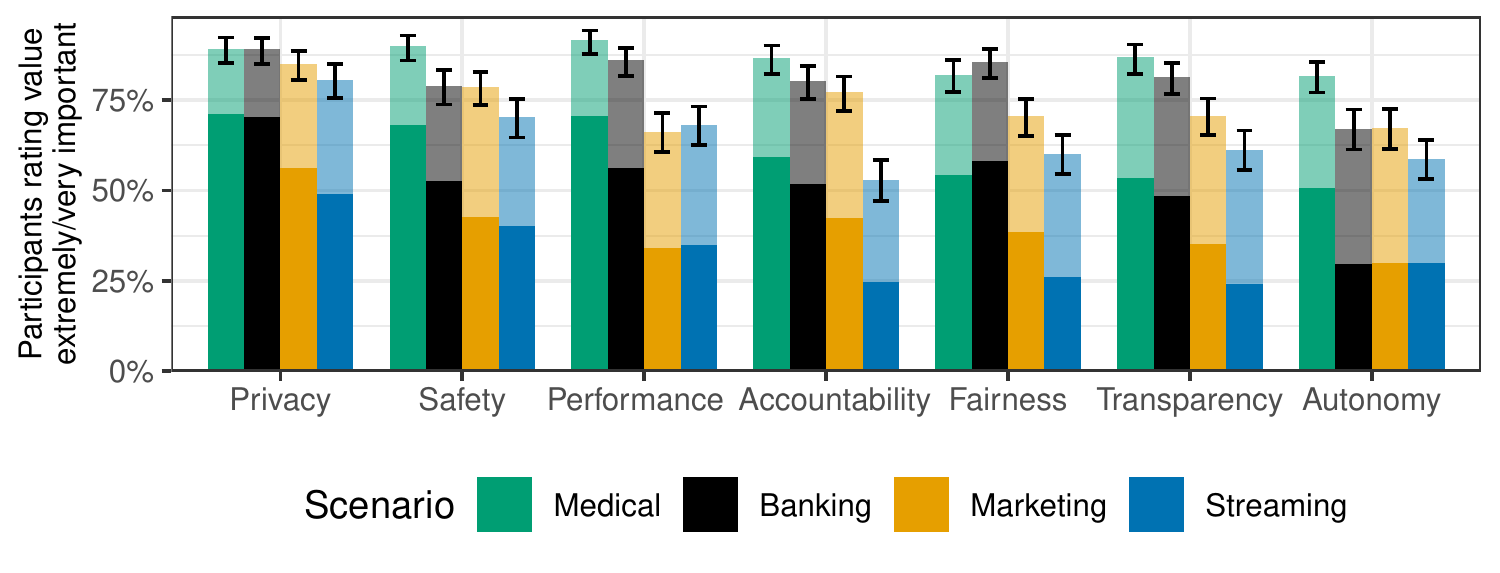}
  \end{center}
\caption{\textbf{Responsible AI values were rated as most important in the medical and banking scenarios.} N=287 to 344 ratings per bar aggregated across samples. The y-axis shows how often respondents evaluated the responsible AI value as very important (light) or extremely important (dark). The perceived importance of other values is dependent on the application context.}
\label{fig:task2-context}
\end{figure*}

In Task 2 participants evaluated how important they considered a value in the context of a specific deployment scenario (Figures~\ref{fig:task2-sample} and~\ref{fig:task2-context}).
The perceived importance of performance, accountability, fairness, and transparency varied significantly across deployment settings. In general responsible AI values were rated as very or extremely important. 
Compared to both the US-census representative and the crowdworker samples, on average, AI practitioners evaluated responsible AI values, and privacy, safety, and performance, in particular, as less important. We also observed significant variation of perceived importance across deployment settings, with responsible AI values being considered most important in the medical context and least important in the streaming context. A more detailed graph showing responses by both sample and scenario is included in the Appendix~\ref{appendix:results}.

\subsection{How values are prioritized when in conflict}

\begin{figure*}
  \begin{center}
    \includegraphics[width=0.8\textwidth, trim=0cm 0.4cm 0cm 0cm]{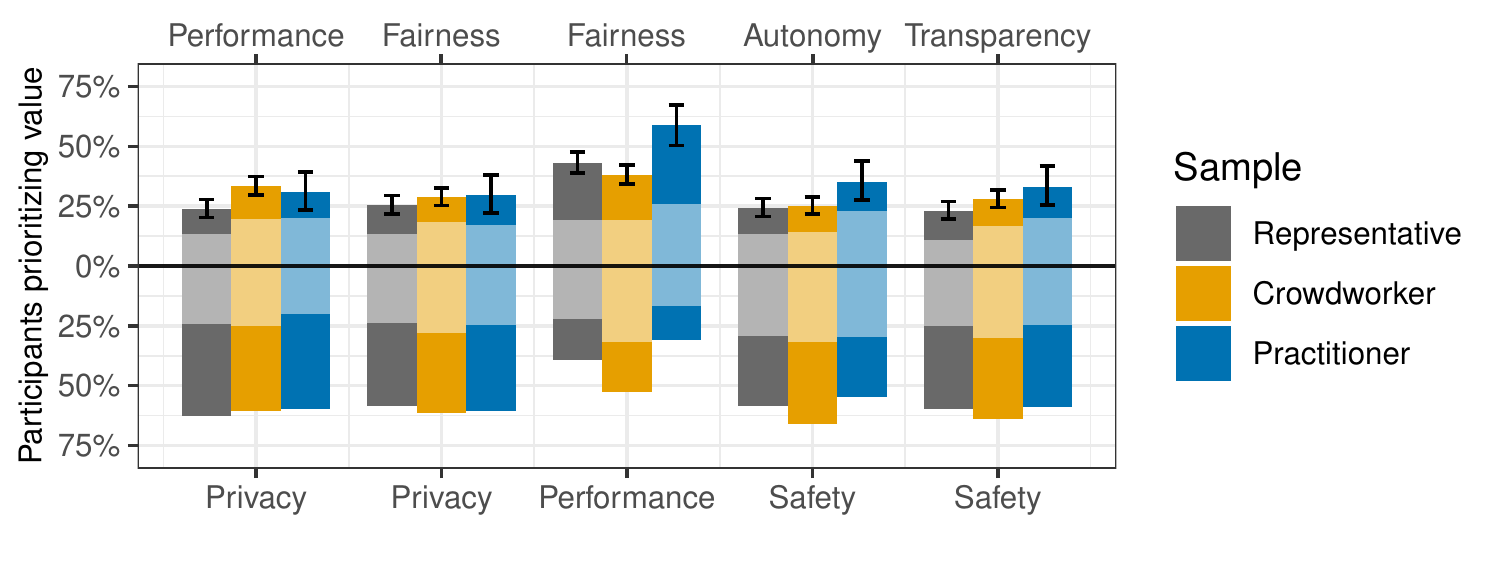}
  \end{center}
\caption{\textbf{Participants across groups prioritized privacy and safety over fairness, but disagreed on the fairness vs. performance tradeoff.} N = 104 to 607 ratings per bar. The conflicting value pairs are shown on the top and bottom, e.g., performance vs. privacy on the left. The proportion of respondents prioritizing the top value is shown to the top and the proportion of respondents prioritizing the bottom value to the bottom. Respondents expressing a strong preferences are shaded in dark, whereas weak preferences are lightly shaded. Undecided respondents are omitted.}
\label{fig:task3-sample}
\end{figure*}

\begin{figure*}
  \begin{center}
    \includegraphics[width=0.8\textwidth, trim=0cm 0.4cm 0cm 0cm]{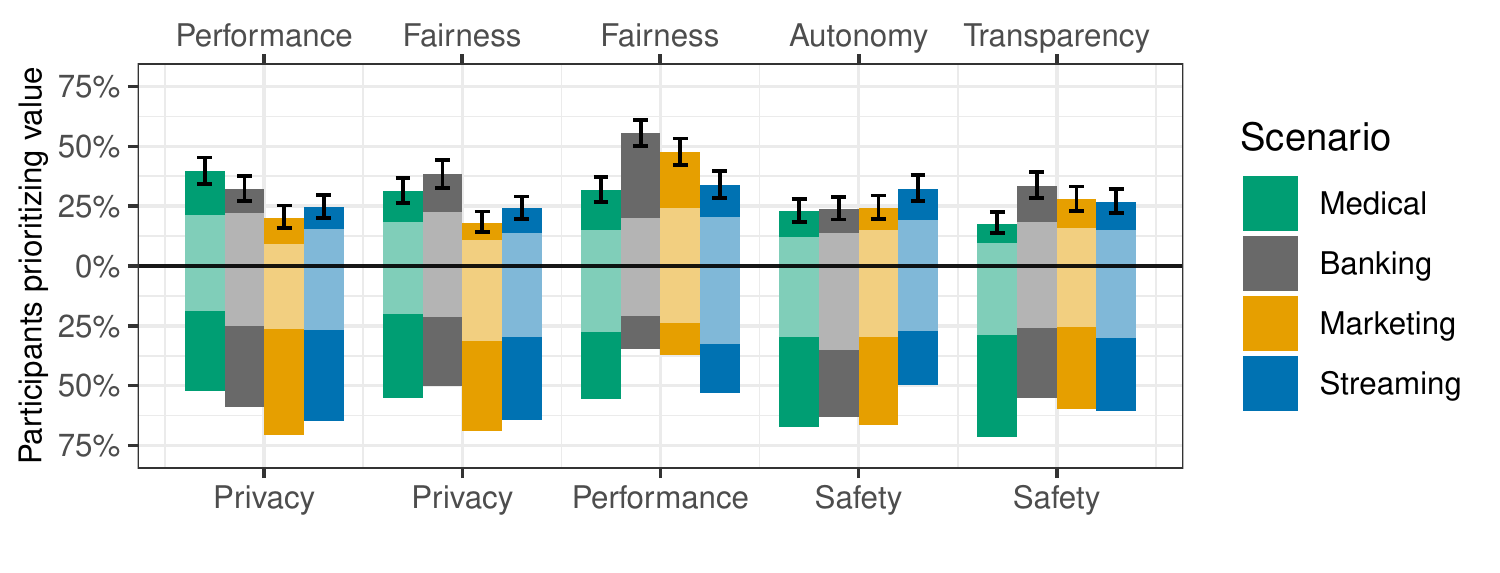}
  \end{center}
\caption{\textbf{Value priorities vary by context, but most participants prioritized privacy and safety across most scenarios.} N = 276 to 341 ratings per bar aggregated across samples. The proportion of respondents prioritizing the top value are shown to the top and the proportion of respondents prioritizing the bottom value to the bottom. Respondents expressing a strong preferences are shaded in dark, weak preferences are lightly shaded.}
\label{fig:task3-context}
\end{figure*}

In Task 3 participants suggested how values should be prioritized when in conflict (Figures~\ref{fig:task3-sample} and~\ref{fig:task3-context}).  
Respondents from all participant samples agreed on prioritizing safety over autonomy and transparency. Across scenarios, a majority of respondents agreed on prioritizing privacy over performance or fairness. Most disagreement was observed when performance and fairness conflicted: 
Participants from the US representative sample were almost equally split in their preferences for fairness versus performance.
Crowdworkers were less likely to prioritize performance and AI practitioners were more likely to prioritize fairness than the US-census representative participants. 

Across scenarios, respondents prioritized privacy over performance and fairness, and safety over autonomy and transparency. Again, 
the performance-fairness trade-off produced most variation: Participants prioritized performance in the medical and streaming scenario, and fairness in the banking and marketing scenario.

\begin{table*}[t] \centering 
  \caption{
  Regression analysis with simple baseline models predicting the value importance ratings based on scenario, sample, and demographic correlates.
  The constant corresponds to a white male respondent from the US-census representative sample assessing a value in the banking scenario. Bold text indicates statistical significance.} 
  \label{table-1} 
\begin{tabular}{@{\extracolsep{5pt}}lccccccc@{}} 
\\[-1.8ex]\hline 
\hline \\[-1.8ex] 
 & \multicolumn{7}{c}{\textit{Dependent variable:}} \\ 
\cline{2-8} 
\\[-1.8ex] & \multicolumn{7}{c}{Value importance rating} \\ 
\\[-1.8ex] & Privacy & Safety & Perform. & Account. & Fairness & Transp. & Autonomy\\ 
\hline \\[-1.8ex] 
 Marketing system & \textbf{-0.051}$^{**}$ & -0.024 & \textbf{-0.14}$^{***}$ & \textbf{-0.046}$^{*}$ & \textbf{-0.10}$^{***}$ & \textbf{-0.08}$^{***}$ & -0.004 \\ 
  Medical system & 0.005 & \textbf{0.06}$^{***}$ & \textbf{0.044}$^{*}$ & 0.030 & -0.031 & 0.029 & \textbf{0.10}$^{***}$ \\ 
  Streaming  system & \textbf{-0.08}$^{***}$ & \textbf{-0.09}$^{***}$ & \textbf{-0.12}$^{***}$ & \textbf{-0.18}$^{***}$ & \textbf{-0.18}$^{***}$ & \textbf{-0.16}$^{***}$ & -0.036 \\ 
  \rule{0pt}{3ex}Crowdworker sample & 0.005 & -0.020 & 0.002 & \textbf{-0.05}$^{***}$ & \textbf{-0.06}$^{***}$ & \textbf{-0.032}$^{*}$ & \textbf{-0.041}$^{*}$ \\ 
 Practitioner sample & \textbf{-0.08}$^{***}$ & \textbf{-0.057}$^{*}$ & -0.052 & \textbf{-0.10}$^{***}$ & \textbf{-0.057}$^{*}$ & \textbf{-0.09}$^{***}$ & 0.005 \\ 
  \rule{0pt}{3ex}Women respondents & \textbf{0.04}$^{***}$ & \textbf{0.039}$^{**}$ & \textbf{0.05}$^{***}$ & \textbf{0.04}$^{*}$ & \textbf{0.05}$^{***}$ & \textbf{0.028}$^{*}$ & \textbf{0.07}$^{***}$ \\ 
  Gender-diverse resp. & -0.042 & -0.031 & -0.046 & -0.017 & 0.086 & -0.010 & 0.041 \\ 
  Black respondents & \textbf{0.046}$^{*}$ & \textbf{0.052}$^{*}$ & \textbf{0.062}$^{**}$ & 0.006 & \textbf{0.08}$^{***}$ & 0.019 & 0.031 \\ 
  Hispanic respondents & 0.024 & 0.042 & 0.044 & 0.034 & 0.020 & 0.021 & 0.023 \\ 
  Asian respondents & 0.001 & -0.025 & -0.017 & -0.018 & -0.031 & \textbf{-0.057}$^{*}$ & -0.007 \\ 
  \rule{0pt}{3ex}Age & 0.001 & -0.0001 & -0.001 & -0.0004 & -0.001 & -0.0003 & 0.001 \\ 
  Education & -0.020 & -0.047 & -0.026 & 0.009 & 0.007 & 0.012 & 0.013 \\ 
  Political leaning & \textbf{0.060}$^{*}$ & 0.011 & 0.027 & 0.023 & \textbf{0.056}$^{*}$ & 0.049 & 0.006 \\ 
  \rule{0pt}{3ex}Exp. with discrimination & -0.027 & 0.011 & \textbf{-0.057}$^{*}$ & 0.002 & 0.004 & -0.008 & 0.005 \\ 
  Familiarity with ML & -0.037 & -0.007 & 0.005 & -0.028 & -0.016 & -0.011 & 0.009 \\ 
  Familiarity with UX & \textbf{0.046}$^{*}$ & 0.043 & \textbf{0.053}$^{*}$ & 0.034 & \textbf{0.055}$^{*}$ & \textbf{0.057}$^{*}$ & 0.043 \\ 
  Constant & \textbf{0.82}$^{***}$ & \textbf{0.80}$^{***}$ & \textbf{0.83}$^{***}$ & \textbf{0.82}$^{***}$ & \textbf{0.81}$^{***}$ & \textbf{0.77}$^{***}$ & \textbf{0.62}$^{***}$ \\ 
 \hline \\[-1.8ex] 
Observations & 1,246 & 1,246 & 1,246 & 1,246 & 1,246 & 1,246 & 1,246 \\ 
R$^{2}$ & 0.082 & 0.084 & 0.150 & 0.130 & 0.119 & 0.111 & 0.070 \\ 
Adjusted R$^{2}$ & 0.070 & 0.072 & 0.139 & 0.118 & 0.107 & 0.099 & 0.058 \\ 
Residual Std. Error & 0.212 & 0.233 & 0.229 & 0.238 & 0.244 & 0.242 & 0.254 \\ 
F Statistic & 6.84$^{***}$ & 7.05$^{***}$ & 13.51$^{***}$ & 11.42$^{***}$ & 10.33$^{***}$ & 9.58$^{***}$ & 5.81$^{***}$ \\ 
\hline 
\hline \\[-1.8ex] 
\textit{Note:}  & \multicolumn{7}{r}{}$^{*}$p$<$0.05; $^{**}$p$<$0.01; $^{***}$p$<$0.001 \\ 
\end{tabular} 
\end{table*}

\begin{table*}[t] \centering 
  \caption{
  Regression analysis with simple baseline models predicting the value preference ratings based on scenario, sample, and demographic correlates.
  The constant corresponds to a white male respondents from the US-census representative sample recommending a value prioritization in the banking scenario. Bold text indicates statistical significance.} 
  \label{table-2} 
\begin{tabular}{@{\extracolsep{5pt}}lccccc} 
\\[-1.8ex]\hline 
\hline \\[-1.8ex] 
 & \multicolumn{5}{c}{\textit{Dependent variable:}} \\ 
\cline{2-6} 
\\[-1.8ex] & \multicolumn{5}{c}{Value preference rating} \\ 
\\[-1.8ex] & Privacy. vs. & Privacy vs. & Performance vs. & Safety vs. & Safety. vs.\\ 
& performance & fairness & fairness & autonomy & transparency\\ 
\hline \\[-1.8ex] 
 Marketing system & \textbf{0.164}$^{**}$ & \textbf{0.301}$^{***}$ & \textbf{0.113}$^{*}$ & 0.067 & 0.080 \\ 
  Medical system & \textbf{-0.112}$^{*}$ & \textbf{0.113}$^{*}$ & \textbf{0.378}$^{***}$ & 0.078 & \textbf{0.259}$^{***}$ \\ 
  Streaming system & 0.091 & \textbf{0.209}$^{***}$ & \textbf{0.342}$^{***}$ & \textbf{-0.150}$^{**}$ & 0.086 \\ 
  \rule{0pt}{3ex}Crowdworker sample & -0.079 & 0.048 & \textbf{0.152}$^{***}$ & 0.058 & 0.015 \\ 
  Practitioner sample & -0.060 & 0.114 & -0.091 & -0.078 & 0.062 \\ 
  \rule{0pt}{3ex}Women respondents & 0.055 & 0.038 & 0.057 & 0.076 & \textbf{0.113}$^{**}$ \\ 
  Gender-diverse resp. & 0.219 & -0.128 & -0.182 & 0.060 & -0.214 \\ 
  Black respondents & -0.006 & -0.098 & -0.109 & \textbf{0.168}$^{**}$ & -0.029 \\ 
  Hispanic respondents & -0.050 & \textbf{-0.160}$^{*}$ & 0.082 & -0.017 & -0.021 \\ 
  Asian respondents & -0.033 & 0.024 & -0.113 & 0.033 & 0.020 \\ 
  \rule{0pt}{3ex}Age & 0.001 & \textbf{0.003}$^{*}$ & -0.002 & 0.0002 & 0.0002 \\ 
  Education & 0.104 & -0.126 & -0.067 & 0.045 & -0.063 \\ 
  Political leaning & 0.010 & \textbf{-0.191}$^{*}$ & \textbf{-0.226}$^{**}$ & 0.091 & -0.017 \\ 
  \rule{0pt}{3ex}Exp. with discrimination & -0.113 & \textbf{-0.205}$^{**}$ & 0.023 & \textbf{-0.160}$^{*}$ & 0.083 \\ 
  Familiarity with ML & 0.008 & 0.073 & -0.008 & 0.090 & -0.078 \\ 
  Familiarity with UX & -0.093 & 0.124 & 0.026 & 0.048 & 0.016 \\ 
  Constant & \textbf{0.262}$^{*}$ & 0.100 & -0.057 & 0.102 & 0.159 \\ 
 \hline \\[-1.8ex] 
Observations & 1,246 & 1,246 & 1,246 & 1,246 & 1,246 \\ 
R$^{2}$ & 0.032 & 0.056 & 0.080 & 0.038 & 0.031 \\ 
Adjusted R$^{2}$ & 0.019 & 0.043 & 0.068 & 0.025 & 0.018 \\ 
Residual Std. Error & 0.712 & 0.679 & 0.701 & 0.665 & 0.688 \\ 
F Statistic & 2.535$^{***}$ & 
4.526$^{***}$ & 
6.719$^{***}$ & 
3.009$^{***}$ & 
2.459$^{**}$ \\ 
\hline 
\hline \\[-1.8ex] 
\textit{Note:}  & \multicolumn{5}{r}{}$^{*}$p$<$0.05; $^{**}$p$<$0.01; $^{***}$p$<$0.001 \\ 
\end{tabular} 
\end{table*}

\subsection{Demographics and experiential correlates of responsible AI value priority}

To explore how demographic and experiential factors correlate with participants' assessments, we mapped their responses to a 5-Likert scale that preserves the direction of the original scale. 
Treating ordinal scales as interval scales is controversial, but the scales in our study have a unit of measurement with comparable-size intervals and a zero point, so a continuous analysis is meaningful and justifiable~\cite{knapp1990treating}. 
To examine how various demographic, experiential, or contextual factors may explain the variance in respondents' assessments, we used linear regression to build simple baseline models that predict their assessments. 


Table \ref{table-1} shows parameter estimates of linear regression models fitted to predict how important respondents consider a value in a specific scenario. The model constant corresponds to a white man from the US-census representative sample evaluating a responsible AI value in the banking scenario. 
The parameter estimates confirm that the perceived importance of values varies significantly across deployment settings. They also confirm that, compared to the US-census representative sample, AI practitioners evaluated most values as less important. Women and black respondents, on average, evaluated most responsible AI values as more important than other groups. Among the experiential correlates, a self-reported liberal political leaning was associated with a higher valuation of privacy. 
Self-reported experiences with discrimination predicted lower perceived importance of performance but were not statistically significantly correlated with other responsible AI values. While familiarity with ML did not predict different value priorities, respondents reporting to be familiar with UX research evaluated most responsible AI values as more important.

Table \ref{table-2} shows parameter estimates predicting participants' preference in the case of conflicting responsible AI values. Positive coefficients correspond to a preference for the top value. Responses vary significantly by deployment context, but only the response to the fairness-performance trade-off varies by sample.
Women respondents were more likely to prioritize safety over transparency than other groups, and black respondents were more likely to prioritize safety over autonomy. While participants reporting experiences of discrimination were more likely to prioritize fairness over privacy, they were not more likely to prioritize fairness over performance than other groups. Instead, participants with liberal political learning were more likely to prioritize fairness over performance and privacy than other groups. 
Familiarity with ML neither predicted a preference for performance over privacy nor fairness.

Some variables were correlated with each other. For example, the practitioner sample contains fewer women respondents (r=-0.14, p<0.01) and black respondents (r=-0.11, p<0.01), but more educated (r=0.33, p<0.01) and liberal-leaning (r=0.2, p<0.01) respondents. Similarly, liberal-leaning respondents were younger (r=-0.13, p<0.01) and more likely to report experiences with ML (r=0.1, p<0.01) and discrimination (r=0.09, p<0.01). However, a correlation analysis (included in the Appendix~\ref{appendix:correlation}) suggests that no covariates were highly correlated (r>0.7). The variance inflation factor remained below 1.5 across all covariates, indicating little to no multicollinearity issues~\cite{hair2009multivariate}.

\section{Discussion}

\para{AI practitioners’ value priorities for responsible AI differ from those of the general public.} 
Our results empirically corroborate a commonly raised concern: AI practitioners’ value preferences for responsible AI are not representative of the value priorities of the wider US population. 
Compared to a US-census representative public, AI practitioners evaluated responsible AI values as less important in general and emphasized a different set of values.

US-census representative and crowdworker respondents agreed on what values they found most important: safety, privacy, and performance. Practitioners, in comparison, were more likely to prioritize fairness, dignity, and inclusiveness. 

These findings align with prior research finding that different groups have different normative expectations of how AI systems should behave in specific situations~\cite{grgic2018human, pierson2017demographics, awad2018moral, hidalgo2021humans}. Our findings extend prior work by demonstrating how AI practitioners' ethical preferences differ from other groups'. We also show that groups not only differ in their judgment of specific behaviors and technical details, but may disagree on the importance of the very values at the core of responsible AI. 
{\em The disagreement in value priorities highlights the importance of paying attention to who gets to define what constitutes ``ethical'' or ``responsible'' AI}. Responsible AI guidelines~\cite{Jobin_Ienca_Vayena_2019} may emphasize a different set of values depending on who writes them and who is consulted. We hypothesize that consulting populations outside the Western world about their priorities for responsible AI would surface even starker disagreement about the values underlying responsible AI~\cite{schwartz2007basic, kapania2022because}.

\para{What might explain the differences in value priorities between AI practitioners’ and other groups?}
Our results provide limited insight into plausible drivers of differences in values. First, women and black respondents assessed responsible AI as more important than other demographic groups. Their relatively low representation in the AI practitioner sample compared to the representative sample (only 40\% and 2.2\% compared to 52\% and 15\% respectively) explains about 15\% of the lower importance ratings AI practitioners assigned to values in general. Increasing the representation of e.g., women and black researchers in AI~\cite{landivar2013disparities, crawford2016artificial, house2016preparing} may thus result in responsible AI values receiving more attention. 

Another demographic variable that robustly predicted differences in value preferences was respondents' political leaning.  Liberal-leaning respondents were 10\% more likely to select fairness amongst the most important values than conservatives, and were 15.5\% more likely to prioritize fairness in the fairness-performance trade-off. Compared to the representative sample, AI practitioner respondents were substantially more likely to self-identify as liberal-leaning  (52\% compared to 26\%), explaining about 27\% of practitioners' different evaluation of fairness. This result is in line with the broader research on value differences along ideological lines~\cite{braithwaite1998value, wetherell2013discrimination}. It highlights that guidelines for responsible AI need to navigate a polarized value landscape. 

Other demographic and experiential variables, however, were less predictive of how our participants assessed responsible AI values. 
Respondents reporting experience with discrimination were more likely to prioritize fairness over privacy, but did not evaluate fairness as more important than other groups. When asked whether developers should prioritize fairness over performance, participants from minoritized groups and participants reporting experience with discrimination were as undecided as other groups. 
While previous work identified performance as the central value in machine learning research~\cite{Birhane_Kalluri_Card_Agnew_Dotan_Bao_2021}, our results do not suggest that AI practitioners or respondents familiar with machine learning were more likely to value performance. Participants trained in user experience research, however, evaluated responsible AI values more important in general.

\para{Can AI practitioners use crowdsourcing to complement their ethical intuitions in the development process?}
Our findings emphasize the need for bringing in a diversity of perspectives when decisions are made about the development and operationalization of responsible AI. 
Crowdworkers are often the go-to convenience sample, but to what extent could they provide a reliable lens into the values that a broader population expect AI systems to adhere to?

As in prior research~\cite{huff2015these}, we find that the value priorities of crowdworkers largely align with those of the US-census representative sample. 
Our results also show that often a majority of participants agreed on value trade-offs. For example, respondents from all samples prioritized privacy over performance across all deployment scenarios.  The agreement raises the question of whether and when product teams could use such results to e.g., justify prioritizing privacy over performance.

Here, consensus alone may not justify practical requirements within specific contexts of use. 
Rather than providing definite answers, the approach developed in this paper provides ``values levers''~\cite{shilton2013values}: organizational processes that take the implicit work of value judgments in technology development and transform it into an explicit matter of debate and documentation. Empirical data on different groups' preferences can both inform the development process of responsible AI and provide opportunities for critical reflection. Rather than prescribing value priorities, responsible AI guidelines could ask practitioners to justify their choices whenever they go against commonly held value preference.

\subsection{Limitations}
The quantitative approach to value elicitation explored above has its benefits: It allows consulting large and representative samples of stakeholders and integrates well with existing crowdwork infrastructures. At the same time, it needs to be complemented by qualitative, small-n investigations like interviews or focus groups for a comprehensive understanding of value differences across social groups. For example, the current study did not explore how  groups understand or interpret values differently, what other values some groups might have wanted to include, or why it is that e.g. women, on average, rated responsible AI values as more important. 

The results of this survey also should be interpreted with care. No normative ``ought’’ can be derived from a descriptive ``is’’~\cite{musschenga2005empirical}. We cannot conclude that safety ought to be prioritized over autonomy from the observation that the respondents in our samples suggested so. Our results aim to increase the context sensitivity of responsible AI decisions, not to prescribe a specific course of action. Empirical ethical research does not replace ethical reasoning but offers perspectives and critical reflections. 

Finally, knowledge-dependent tensions arise when contrasting the perspectives of experts and laypeople. One may argue that non-expert perspectives lack the technical and organizational insight required to evaluate AI systems. However, as we are focusing on ethical rather than technical questions, non-experts have their own valid and legitimate forms of knowledge~\cite{harding1992rethinking} that experts might not be aware of. 

\section{Conclusion}
Recently published guidelines for responsible AI seem to converge on a set of central values. However, little is known about the values a more representative public would find important for responsible AI. We conducted a survey comparing how US-representative respondents, crowdworkers, and AI practitioners perceive and prioritize responsible AI values. Our findings show that, compared to the general public, AI practitioners find responsible AI values less important and are likely to focus on a different set of values. Our findings underline the need for more diverse ethical judgement to be incorporated into the AI development process. Crowdworkers, who are already involved in the AI development process, resemble the general public in their value priorities and might provide valuable input.

\begin{acks}
We thank our colleagues from across Microsoft who provided insight and expertise that greatly assisted the research.
We are particularly grateful to Su Lin Blodgett, Stephanie Ballard, Michael Madaio, Emery Fine and Kate Crawford for their comments on the research framing and survey design.
\end{acks}

\bibliographystyle{ACM-Reference-Format}
\bibliography{references}

\appendix

\section{Appendix: Supplementary materials}


\subsection{Introduction and task}\label{appendix:intro}
Artificial intelligence (AI) is a set of emerging technologies concerned with building smart systems or machines capable of performing tasks that typically require human intelligence. Besides technical challenges, building AI systems involves complex decision-making on what the system should or should not do. In this survey, we will ask you to assess the importance of ethical principles for four AI systems.

\subsection{Value Description and Question Framing}
\label{appendix:RAI-values}
{\footnotesize \def\arraystretch{1}
\begin{tabular}{@{}p{0.15\columnwidth}|p{0.8\columnwidth}@{}}
{\bf RAI value} & {\bf Description} \\\hline
Transparency & A transparent AI system produces decisions that people can understand. Developers of transparent AI systems ensure, as far as possible, that users can get insight into why and how a system made a decision or inference. How important is it that the system is transparent? \\\hline
Fairness & A fair AI system treats all people equally. Developers of fair AI systems ensure, as far as possible, that the system does not reinforce biases or stereotypes. A fair system works equally well for everyone independent of their race, gender, sexual orientation, and ability. How important is it that the system is fair? \\\hline
Safety & A safe AI system performs reliably and safely. Developers of safe AI systems implement strong safety measures. They anticipate and mitigate, as far as possible, physical, emotional, and psychological harms that the system might cause. How important is it that the system is safe? \\\hline
Accountability & An accountable AI system has clear attributions of responsibilities and liability. Developers and operators of accountable AI systems are, as far as possible, held responsible for their impacts. An accountable system also implements mechanisms for appeal and recourse. How important is it that the system is accountable? \\\hline
Privacy & An AI system that respects people's privacy implements strong privacy safeguards. Developers of privacy-preserving AI systems minimize, as far as possible, the collection of sensitive data and ensure that the AI system provides notice and asks for consent. How important is it that the system respects people's privacy? \\\hline
Autonomy & An AI system that respects people's autonomy avoids reducing their agency. Developers of autonomy-preserving AI systems ensure, as far as possible, that the system provides choices to people and preserves or increases their control over their lives. How important is it that the system respects people's autonomy? \\\hline
Performance & A high-performing AI system consistently produces good predictions, inferences or answers. Developers of high-performing AI systems ensure, as far as possible, that the system’s results are useful, accurate and produced with minimal delay. How important is it that the system performs well? \\\hline
\end{tabular}}

\subsection{Value conflict framing}
\label{appendix:conflicts}
{\footnotesize \def\arraystretch{0.9}
\begin{tabular}{@{}p{0.15\columnwidth}|p{0.8\columnwidth}@{}}
{\bf Value pair} & {\bf Description} \\\hline
Fairness vs. performance & The developers realize that making the system treat all people equally (ensuring fairness) may make the system’s predictions less accurate (reducing performance). Should they prioritize fairness or performance? \\\hline
Fairness vs. performance (reverse) & The developers realize that making the system’s predictions possibly accurate (ensuring performance) may mean that the system cannot treat all people equally (reducing fairness). Should they prioritize performance or fairness? \\\hline
Fairness vs. privacy & The developers realize that making the system treat all people equally (ensuring fairness) may require the collection of additional sensitive data (reducing privacy). Should they prioritize fairness or privacy? \\\hline
Fairness vs. privacy (reverse) & The developers realize that minimizing the collection of sensitive data (ensuring privacy) may mean that the system cannot treat all people equally (reducing fairness). Should they prioritize privacy or fairness? \\\hline
Privacy vs. performance & The developers realize that minimizing the collection of sensitive data (ensuring privacy) may make the system’s predictions less accurate (reducing performance). Should they prioritize privacy or performance? \\\hline
Privacy vs. performance (reverse) & The developers realize that making the system’s predictions possibly accurate (ensuring performance) may require the collection of additional sensitive data (reducing privacy). Should they prioritize performance or privacy? \\\hline
Safety vs. autonomy & The developers realize that mitigating risks and potential harms (ensuring safety) may require limiting people’s choices and control (reducing autonomy). Should they prioritize safety or people's autonomy? \\\hline 
Safety vs. autonomy (reverse) & The developers realize that giving people choices and control (ensuring autonomy) may introduce additional risks and potential harms (reducing safety). Should they prioritize people's autonomy or safety? \\\hline 
Safety vs. transparency & The developers realize that mitigating risks and potential harms (ensuring safety) may require to keep the system’s decision process opaque (reducing transparency). Should they prioritize safety or  transparency? \\\hline 
Safety vs. transparency (reverse) & The developers realize that revealing the system’s decision process (ensuring transparency) may introduce additional risks and potential harms (reducing safety). Should they prioritize transparency or safety? \\\hline
\end{tabular}}

\subsection{Application scenario framing}
\label{appendix:scenarios}
{\footnotesize \def\arraystretch{0.9}
\begin{tabular}{@{}p{0.15\columnwidth}|p{0.8\columnwidth}@{}}
{\bf Scenario} & {\bf Description} \\\hline
Banking & A bank uses an AI system that scans loan applicants' data to predict whether they are likely to repay a loan. Thousands of loan applications are automatically rejected based on the output of this AI system. \\\hline
Medical & A medical clinic uses an AI system that scans patients' medical records to predict whether a patient has a particular disease. Thousands of patients' treatment plans are automatically adjusted based on the output of this AI system. \\\hline
Marketing & A marketing company uses an AI system that scans the data of web users to predict which advertisements they will respond to. Thousands of advertisements are automatically shown to users based on the output of this AI system. \\\hline 
Streaming & A video streaming company uses an AI system that scans users' data to predict which other movies they would enjoy seeing. A list of recommended movies is automatically shown to thousands of users based on the output of this AI system.\\\hline
\end{tabular}}

\subsection{Detailed result graphs}
Please refer to Figures 7 and 8.

\subsection{Covariate correlation analysis}
\label{appendix:correlation}
\begin{table*}[ht]
\centering
{\footnotesize \def\arraystretch{0.9}
\begin{tabular}{rp{2.75em}p{2.75em}p{2.75em}p{2.75em}p{2.75em}p{2.75em}p{2.75em}p{2.75em}p{2.75em}p{2.75em}p{2.75em}p{2.75em}}
  \hline
   & Crowd-workers & Practi-tioners & Women resp. & Diverse resp. & Black resp. & Hispanic resp. & Asian resp. & Age & Edu-cation & Pol. lean. & Dis-crimin. & Fam. ML \\ 
  \hline
  Women respondents &  0.01     & -0.14** &  &  &  &  &  &  &  &  &  &  \\ 
  Gender-diverse resp. &  0.07*    & -0.01     & -0.15** &  &  &  &  &  &  &  &  &  \\ 
  Black respondents &  0.01     & -0.11** &  0.05     & -0.05     &  &  &  &  &  &  &  &  \\ 
  Hispanic respondents & -0.01     & -0.08**   & -0.02     &  0.05     & -0.02     &  &  &  &  &  &  &  \\ 
  Asian respondents &  0.09**   &  0.06*    & -0.05     &  0.04     & -0.09**   & -0.07*    &  &  &  &  &  &  \\ 
  Age & -0.24** & -0.14** &  0.05     & -0.10**  & -0.04     & -0.07*    & -0.16** &  &  &  &  &  \\ 
  Education &  0.06*    &  0.33** & -0.09**   &  0.00     & -0.09**   & -0.07*    &  0.12** &  0.05     &  &  &  &  \\ 
  Political leaning &  0.03     &  0.20** & -0.05     &  0.15** & -0.01     &  0.04     &  0.05     & -0.13** &  0.17** &  &  &  \\ 
  Exp. w. discrimination &  0.01     &  0.14** &  0.04     &  0.15** &  0.16** &  0.05     &  0.07*    & -0.14** &  0.08**   &  0.09**   &  &  \\ 
  Familiarity with ML & -0.02     &  0.23** & -0.08**   &  0.02     & -0.01     & -0.01     &  0.13** & -0.16** &  0.22** &  0.10**  &  0.18** &  \\ 
  Familiarity with UX &  0.10**  &  0.05     & -0.01     & -0.01     &  0.06*    & -0.02     &  0.09**  & -0.13** &  0.18** &  0.01     &  0.19** &  0.42** \\ 
   \hline
\textit{}  & \multicolumn{12}{r}{Note: $^{*}$p$<$0.05; $^{**}$p$<$0.01} \\ 
\end{tabular}}
\end{table*}

\label{appendix:results}
\begin{figure*}[t]
  \begin{center}
    \includegraphics[width=0.9\textwidth, trim=0cm 0.4cm 0cm 0cm]{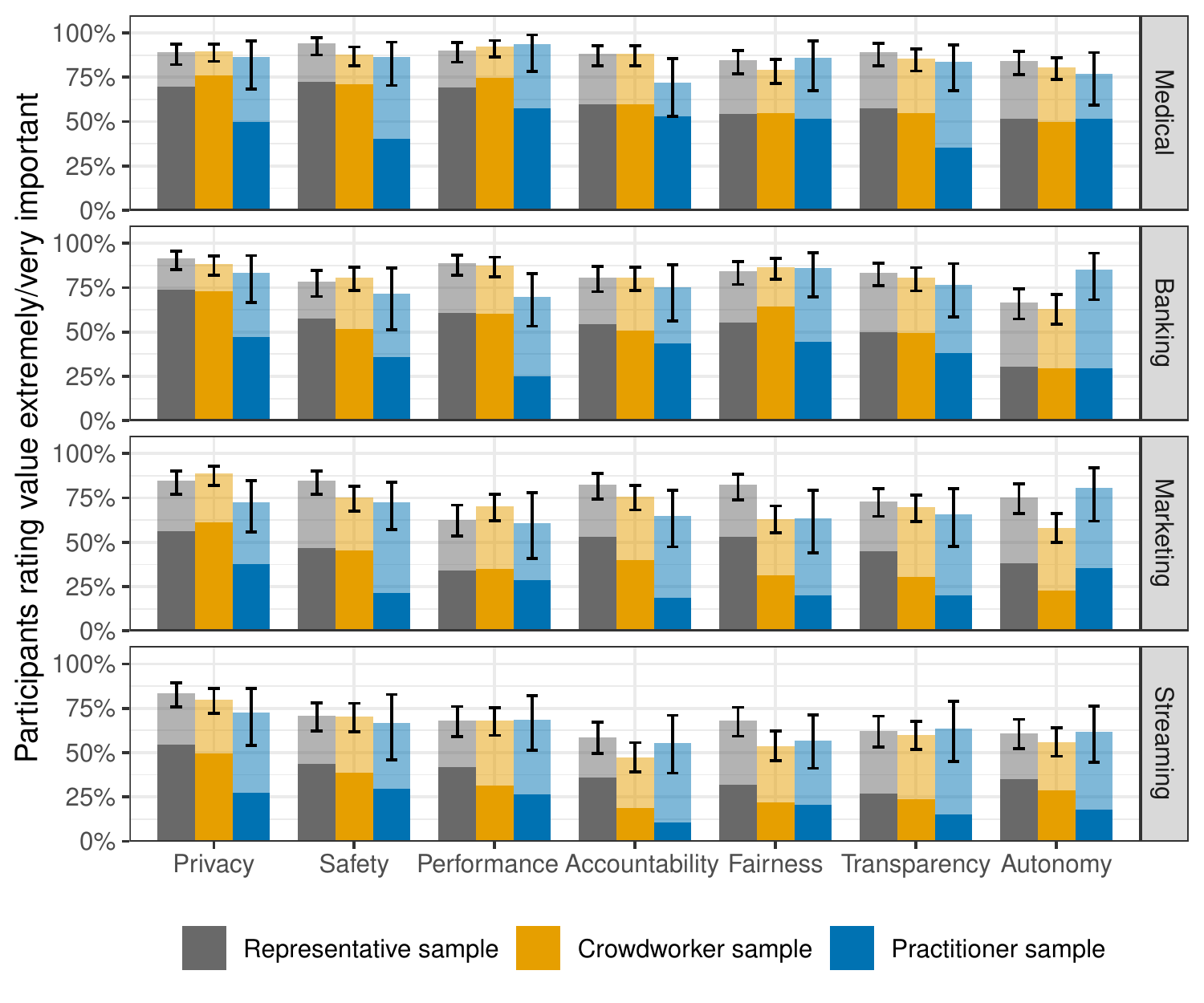}
  \end{center}
\caption{\textbf{The perceived importance of values across deployment scenarios.} N=28 to 171 ratings per bar. The x-axis shows the assessed responsible AI values and the y-axis indicates how often respondents evaluated the responsible AI value as very important (light) or extremely important (dark).}
\label{fig:supplement-task2}
\end{figure*}

\begin{figure*}
  \begin{center}
    \includegraphics[width=0.8\textwidth, trim=0cm 0.4cm 0cm 0cm]{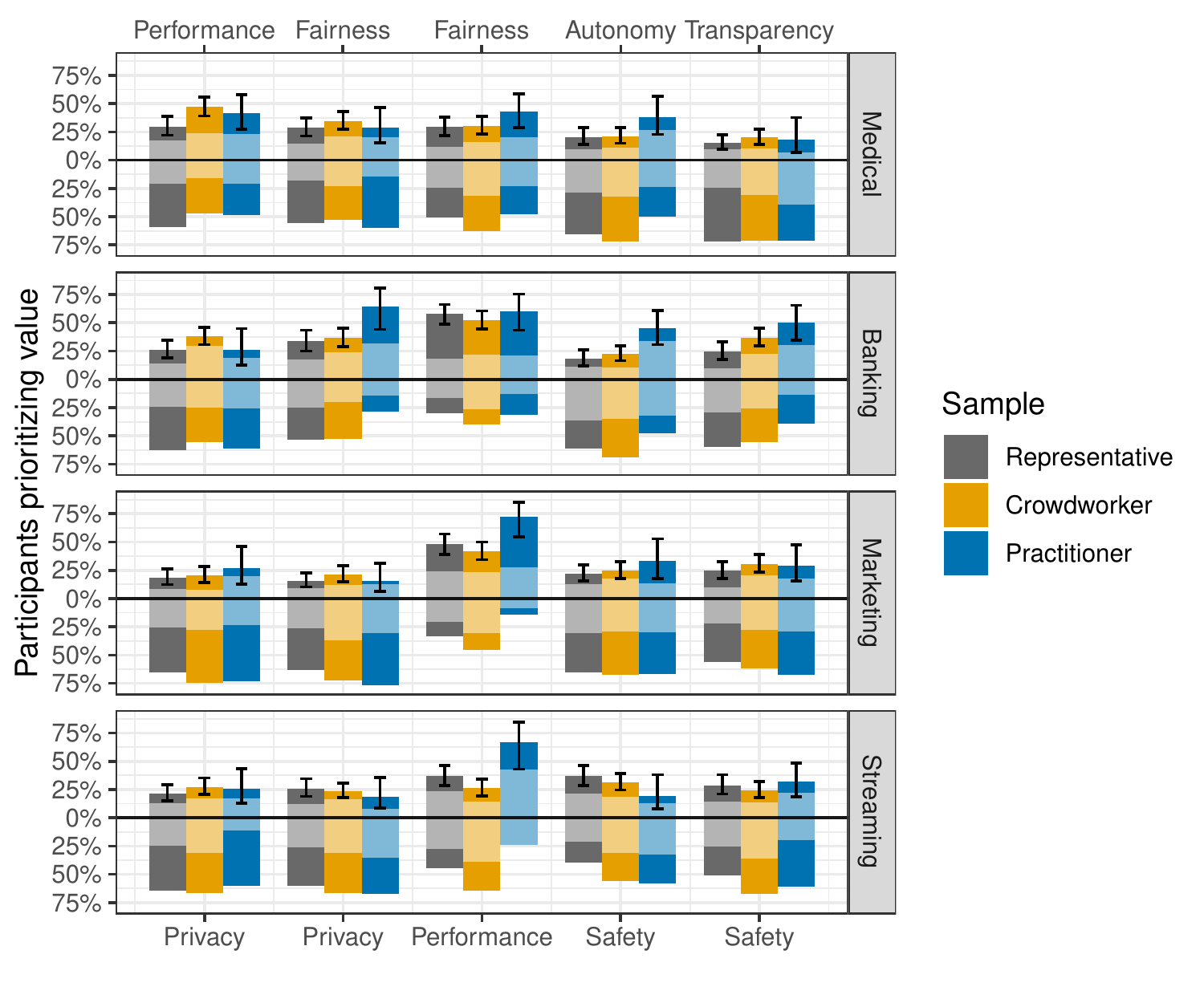}
  \end{center}
\caption{\textbf{How values are prioritized in different deployment settings.} N = 28 to 173 ratings per bar. The conflicting value pairs are shown on the top and bottom, e.g., privacy vs. performance on left. Respondents prioritizing the top value are shown to the top and responses prioritizing the bottom value to the bottom. Respondents expressing a strong preferences are shaded in dark, whereas weak preferences are lightly shaded. Undecided respondents are omitted.}
\label{fig:supplement-task3}
\end{figure*}

\end{document}